\documentclass[aps,prd,preprint,eqsecnum,tightenlines,floats,
groupedaddress,showpacs]{revtex4}
\usepackage{amssymb}
\usepackage{graphicx}
\usepackage{subfigure}
\usepackage{dcolumn}

\begin{document}

\date{January 26, 2026}

\title{Supermassive black hole formation in the initial collapse\\
of axion dark matter}

\author{Pierre Sikivie and Yuxin Zhao}
\affiliation{Department of Physics, University of Florida,
Gainesville, FL 32611, USA}

\vskip 1cm

\begin{abstract}

Axion dark matter thermalizes by gravitational 
self-interactions and forms a Bose-Einstein condensate. 
We show that the rethermalization of the axion fluid 
during the initial collapse of large scale overdensities
near cosmic dawn transports angular momentum outward 
sufficiently fast that black holes form with masses 
ranging from approximately $10^5$ to a few times 
$10^{10}M_\odot$.  This conclusion holds for QCD 
axions and for axion-like particles of mass larger 
than $10^{-16}$ eV/$c^2$.

\end{abstract}


\maketitle

It has long been established \cite{Korm} that most 
galaxies have supermassive black holes at their centers
with masses ranging from approximately $10^5 M_\odot$ 
to a few times $10^{10} M_\odot$. The event horizons 
of the black holes at the center of the Milky Way and 
the center of the large galaxy M87 have been imaged 
using very long baseline interferometry \cite{EHT}. 
Active galactic nuclei (AGN) are understood to be
powered by accretion onto supermassive black 
holes~\cite{AGN}. Recently, low frequency gravitational
waves consistent with emission from supermassive black 
hole mergers have been detected by pulsar 
timing arrays \cite{PTA}.

How the supermassive black holes form has been an 
enduring puzzle. The main impediment to their 
formation is conservation of angular momentum. If, 
for example, a $10^8 M_\odot$ black hole condenses 
out of a region of density, say $10^{-23}$ gr/cc, which 
is a value typical of galactic disks, the material 
forming the black hole must shrink in all directions 
by eight orders of magnitude. Angular momentum 
conservation makes this difficult by introducing a 
distance of closest approach to the black hole. 
Angular momentum can be transported outward if the 
material falling toward the black hole has viscosity 
but in that case the material heats up and acquires 
pressure opposing its compression. See Ref. \cite{bhfrev} 
for a review of the issues involved in supermassive black 
hole formation.

To sidestep the problems that arise when only 
conventional physics is involved, it has been 
proposed that the supermassive black holes form 
as a result of the gravothermal collapse of 
overdensities of dark matter with very strong 
self-interactions \cite{SS} or by accretion 
onto dark stars, i.e. stars that are powered by dark 
matter annihilation \cite{KF}. In such scenarios, 
seed black holes form that have masses of order 
$10^5 \text{--} 10^6~M_\odot$. Larger supermassive
black holes with mass $10^9 \text{--} 10^{10}~M_\odot$
are then supposed to be the result of accretion 
onto, and mergers of, the seed black holes. 
These proposals have been challenged by the 
recent discovery, using the James Webb Space 
Telescope, of powerful AGN near cosmic dawn, i.e. 
at redshifts $z \sim 10$ \cite{JWST}. There is 
very little time to grow the black holes powering 
the AGN observed at $z \sim 10$. Finally, one 
may contemplate the possibility that the 
supermassive black holes are primordial in 
nature, i.e. that they formed long before 
cosmic dawn. This proposal runs afoul of 
constraints from cosmic microwave background 
observations, although a recent paper \cite{HoSt} 
indicates how it may still be viable.

The purpose of our paper is to show that supermassive
black holes form naturally near cosmic dawn if the 
dark matter is axions or axion-like particles. No
additional assumptions are required. The crucial 
step is to recognize that cold dark matter axions 
thermalize by their gravitational self-interactions 
\cite{CABEC,Erken}. Indeed, as explained in more 
detail below, their density fluctuations are 
generically large $(\delta \rho \sim \rho)$ and 
correlated over long distances. When the axions 
thermalize they form a Bose-Einstein condensate, 
meaning that most axions go to the lowest energy 
state available to them through the thermalizing 
interactions. When an axion overdensity collapses 
near cosmic dawn, the gravitational self-interactions 
among the axions produce a long range viscosity 
that causes outward transport of angular momentum. 
The heat produced in the axion case flows into 
the thermal distribution accompanying the condensate 
whereas the condensate, containing most of the 
axions, stays in the lowest energy particle state 
available. That state is one of rigid rotation 
where most of the angular momentum resides far 
from the central overdensity which may then 
perhaps collapse into a black hole.

First let us indicate how axions differ from
other cold dark matter candidates, such as weakly
interacting massive particles (WIMPs) and sterile 
neutrinos, which we will refer to collectively as 
``ordinary cold dark matter (CDM)''. Before density 
perturbations go non-linear and multi-streaming begins, 
in particular during the recombination era, the 
state of ordinary CDM is customarily given by 
a mass density $\rho(\vec{x},t)$ and a velocity 
field $\vec{v}(\vec{x},t)$. In linear order of 
perturbation theory, axions behave in the same 
way as ordinary CDM on length scales longer 
than their Jeans length \cite{Zel}. When the 
axion fluid has mass density $\rho(\vec{x},t)$ 
and velocity field $\vec{v}(\vec{x},t)$, all 
axions are in the particle state of wavefunction
\begin{equation}
  \Psi(\vec{x},t) =
  \sqrt{\rho(\vec{x},t) \over N m} 
e^{i \beta (\vec{x},t)} \,,
\label{psi}
\end{equation}
where $m$ is the axion mass, $N$ is the number 
of axions and $\beta(\vec{x},t)$ is such that 
\begin{equation}
  \vec{v}(\vec{x},t) = 
  {\hbar \over m} \vec{\nabla} \beta(\vec{x},t) \,.
\end{equation}
For both ordinary CDM
and axions, the above description is exact only 
when the dark matter has zero velocity dispersion.
In reality both axions and ordinary CDM have 
a finite velocity dispersion. In the case of 
ordinary CDM this distinction is irrelevant 
because ordinary CDM is non-degenerate. But 
it is important in the case of axions because 
CDM axions form a degenerate Bose gas \cite{Erken}. 
WIMPs of mass 100~GeV have primordial velocity 
dispersion $\delta v_W \sim 10^{-12}~c~a(t_0)/a(t)$ 
and hence correlation length 
$\ell_W \sim 2~\mathrm{\mu m}~a(t)/a(t_0)$, where 
$a(t)$ is the scale factor, $t$ is cosmic time 
and $t_0$ is the present age of the universe. 
Sterile neutrinos with mass a few keV have 
primordial velocity dispersion 
$\delta v_\nu \sim 10^{-8}~c~a(t_0)/a(t)$
and hence $\ell_\nu \sim$ cm $a(t)/a(t_0)$. The 
correlation lengths of ordinary CDM are far too 
short to be relevant in large scale structure
formation. Axions are different.

The correlation length and velocity dispersion 
of CDM axions are set by the horizon 
when the axion mass turns on during the QCD phase 
transition: 
\begin{equation}
  \ell(t) \sim c t_1 {a(t) \over a(t_1)}\,,\quad
  \delta v(t) \sim 
  {\hbar \over m c t_1}{a(t_1) \over a(t)}
\label{veldis}
\end{equation}
where $t_1 \simeq 4 \times 10^{-7}\,{\rm s}
({\mu{\rm eV} \over mc^2})^{1 \over 3}$ is the 
time when axion field oscillations begin 
\cite{axdm}. The thermal relaxation rate 
of the axion fluid is of order \cite{CABEC,Erken}
\begin{equation}
\Gamma(t) \sim 4 \pi G \rho(t) m \ell(t)^2/\hbar\, .
\label{relax}
\end{equation}
This result can be understood by noting that the 
axion fluid has density fluctuations
$\delta \rho = \rho$ correlated over distances 
of order $\ell$; see Appendix A. The density 
fluctuations source gravitational fields 
$\delta g \sim 4 \pi G \rho \ell$. Since the fluid 
momentum dispersion is $\delta p \sim \hbar/\ell$, 
the fluctuating gravitational forces $m \delta g$ 
modify the momentum distribution entirely in a time
$\tau \sim \delta p/m \delta g$. The thermal 
relaxation rate $\Gamma$ is the inverse of 
$\tau$.  The reader may be surprised that the 
relaxation rate given in Eq.~(\ref{relax}) is of 
order $G$ rather than of order $G^2$.  Ref. \cite{PoS} 
discusses in detail why the familiar formula for 
the relaxation rate, which scales as $G^2$, does 
not apply in the situation discussed here.  
The order $G^2$ formula would overestimate the 
relaxation rate by many orders of magnitude.  It
is inapplicable in the situation of interest because 
it relies on Fermi's Golden Rule which assumes that 
each individual axion scattering happens separately 
from all the other axion scatterings, so that energy 
is conserved in each scattering separately.  This 
assumption fails here because the scatterings happen 
so fast that they overlap in time. In spite of its
unfamiliarity, we assume Eq.~(\ref{relax}) is 
correct, both for the reasons given in 
Ref.~\cite{Erken} and because it follows, as
explained above, from the fact that the axion 
fluid has density fluctuations $\delta \rho = \rho$ 
correlated over distances of order $\ell$.

One may attempt a classical field theory 
description of the thermalization 
of the cold axion fluid.  In such a 
description, all axions are in a single 
state whose wavefunction satisfies the 
Schr\"odinger-Poisson equations.  The 
wavefunction is a superposition of 
many waves with random phases.  In that 
description the axion fluid also has 
$\delta \rho \simeq \rho$ fluctuations 
correlated over distances of order the 
inverse of the wavevector dispersion, 
and the resulting fluctuations in the 
gravitational field also thermalize the 
axion fluid.  The thermalization rate is 
the same rate, Eq.~(\ref{relax}), as in 
the quantum field theory.  However, the 
outcome of the thermalization process 
is different.  The outcome of thermalization 
in classical field theory is energy 
equipartition among all the field modes.  
The outcome of thermalization in the quantum 
field theory is Bose-Einstein condensation.  
Bose-Einstein condensation plays an essential 
role in the process we describe because the 
lowest energy state for given angular 
momentum is one of rigid rotation where 
most of the angular momentum has moved 
toward the periphery.

$\Gamma(t)$ exceeds the Hubble rate $H(t)$ at 
some time well before matter-radiation equality
\cite{CABEC}.  At that time all conditions for 
Bose-Einstein condensation are satisfied, and 
almost all axions go to the lowest energy particle 
state available to them through the thermalizing 
interactions. The axion fluid correlation length 
consequently grows to be of order the horizon at 
the time. In linear order of perturbation theory, 
the axion condensate evolves as ordinary CDM. 
However, in higher orders and in particular when 
it acquires angular momentum by tidal torquing, 
the condensate can and will lower its energy by 
acquiring vorticity whereas ordinary CDM remains 
vorticity free \cite{Banik}.

Consider a large overdensity of dark matter that 
dominates its environment and is about to collapse 
near cosmic dawn. We ignore at first all density 
perturbations on scales smaller than that of the 
overdensity itself, effectively smoothing it. Its 
smoothed density field has the general form
\begin{equation}
  \rho(\vec{r},t) = \rho(\vec{0},t)
  \left[1 - \left({x_1 \over R_1(t)}\right)^2
  - \left({x_2 \over R_2(t)}\right)^2 - 
  \left({x_3 \over R_3(t)}\right)^2\right] \,,
\label{inden}
\end{equation}
where $(x_1,x_2,x_3)$ are appropriately chosen 
Cartesian coordinates centered on the peak, and 
the $R_i\,(i = 1, 2, 3)$ give the peak's extent in 
the three spatial directions. Ref.~\cite{Bard} 
derived the properties of such peaks in Gaussian 
random fields, e.g. the number of peaks per unit 
volume of a given size, and probability 
distributions for $R_i$. 

Fig.~1 shows the evolution of such an overdensity 
in a 2-dimensional cut $(x,\dot{x})$ of its 
6-dimensional phase-space. $x$ is the spatial 
coordinate along an arbitrary direction through the 
overdensity and $\dot{x}$ the corresponding velocity. 
Curve~(b) shows the distribution of particles at 
time $t_{\rm in}$ defined as the time when the 
central part of the overdensity is at turnaround, 
i.e. when it stops expanding and is about to contract. 
The central density at that time 
$\rho(0, t_\mathrm{in})$ is related to $t_\mathrm{in}$ 
by
\begin{equation}
  t_\mathrm{in} = \frac{\pi}{2}
  \sqrt{\frac{3}{8\pi G \rho(0,t_\mathrm{in})}} \,.
\label{eq:t_in}
\end{equation}
The central part of the overdensity collapses at time 
$t_\mathrm{coll} \simeq 2 t_\mathrm{in}$. Curve~(d) 
shows the distribution of particles at time 
$t_{\rm coll}$. At collapse time the density is 
very large near the center but no black hole forms in
case of ordinary CDM because the particles have 
acquired angular momentum through tidal torquing. 

The amount of angular momentum acquired by galaxies 
through tidal torquing is commonly given by a 
dimensionless number $\lambda$ called the galactic 
spin parameter \cite{Peeb}. Spin parameters in the 
range $0.01 \lesssim \lambda \lesssim 0.18$ are 
predicted \cite{Efst} and found to be consistent 
with the amount of angular momentum baryons are 
observed to carry in galaxies \cite{Hern}. We take 
this range to be a guide to the amount of angular 
momentum that the overdensity under consideration 
acquired through tidal torquing. If the dark matter 
is ordinary CDM, the angular momentum introduces an 
average distance of closest approach to the center 
of the overdensity of order $\lambda^2 R$, far too 
large for a black hole to form. 

If axions are the dark matter, the axions at 
a distance $r$ from the center of the overdensity 
thermalize at the rate given in Eq.~(\ref{relax})
with $\ell \sim r$ since the gravitational fields
due to the axion fluid outside the region
of radius $r$ do not help the thermalization 
within that region \cite{Erken}. Since
$\rho(\vec{r},t)$ exceeds the average cosmological
energy density $\bar{\rho}(t) = 1/6 \pi G t^2$, 
\begin{equation}
  \Gamma(t) \gtrsim 2 \times 10^{16} H(t)
  \Bigg({mc^2 \over \mu{\rm eV}}\Bigg) 
  \Bigg({r \over 10^{22}~{\rm cm}}\Bigg)
  \Bigg({220~{\rm Myr} \over t}\Bigg) \,,
\label{exc}
\end{equation}
where $H(t) = {2 \over 3 t}$ near cosmic dawn, 
the thermalization rate is therefore very large 
compared to the dynamical evolution rate at time 
$t_{\rm in}$. One may readily verify that it 
remains much larger than the dynamical evolution rate
during the collapse. However, thermalization does not 
suffice to justify the angular momentum transport 
necessary for black hole formation. Since the 
axions can only change their momenta by an amount 
$\delta p \sim \hbar / \ell$ in a time $\tau
\sim 1/\Gamma$, they can only change their specific
angular momentum $L$, i.e. their angular momentum 
per unit mass, by an amount $\delta L \sim \hbar/m$ 
in that time. Hence, 
\begin{equation}
  \dot{L}_{\rm max}(r,t) \sim 4 \pi G \rho(r,t) r^2
\label{maxLdot}
\end{equation}
is the maximum rate at which axions at radius $r$ 
can gain or lose specific angular momentum.

The initial overdensity is not spherically 
symmetric, and hence its collapse is not 
isotropic. Instead, its asphericity grows
during the collapse \cite{anis} as the overdensity 
tends to become a pancake or spindle. In our treatment 
below, we ignore the fact that the infall is highly 
anisotropic because anisotropy only produces 
velocities in the angular (i.e. non-radial) directions 
that are at most of order $\sqrt{G M/r}$ where $M$ 
is the mass enclosed by the shell of radius $r$. 
Since this is always much less than $c$ during 
the infall, the angular velocities do not greatly 
affect whether a shell falls within the black hole 
horizon of the mass $M$ it contains. Although we 
ignore the effect of anisotropy on the infall 
motion we do not ignore its effect on the tidal 
torquing experienced by the infalling overdensity. 
Tidal torquing is important because it produces
angular momentum and hence a minimum radius 
\begin{equation}
  r_{\rm min} = {L^2 \over 2 G M} + {\cal O}(L^4)
\label{minrad}
\end{equation}
for each shell. 

In case of isotropic radial infall the radius 
of the shell containing mass $M$ at time $t$ 
is given by the parametric equations
\begin{eqnarray}
  r(M,t) &=& r_0(M) \sin^2 \sigma
  \nonumber\\
  t &=& \sqrt{r_0(M)^3 \over 2 G M} 
  [\sigma - {1 \over 2} \sin(2 \sigma)] \,,
\label{radinf}
\end{eqnarray}
where $r_0(M)$ is the shell's turnaround radius. 
The shell's turnaround time is 
\begin{equation}
  t_0(M) = {\pi \over 2} \sqrt{r_0(M)^3 \over 2 G M} \,.
\label{tat}
\end{equation}
The axion mass density at shell $M$ is
\begin{equation}
\rho(M,t) = {1 \over 4 \pi r(M,t)^2} {d M \over dr}(M,t) \,. 
\label{dens}
\end{equation}
By using Eqs.~(\ref{radinf}) to describe the collapse 
of an axion dark matter overdensity four approximations
are made in addition to ignoring the fact that the 
infall is highly anisotropic. First, we ignore all 
forms or matter and energy (baryons, photons, 
neutrinos, dark energy) other than the axionic dark 
matter. Second, we use Newtonian gravity and Newton's 
laws of motion to describe the infall. Relativistic 
corrections are unimportant until a shell approaches 
the black hole horizon and are unlikely to change 
the final black hole mass by more than a factor two, 
or so, which is within the uncertainty that we 
tolerate throughout. Third, although we keep track 
of the specific angular momentum $L(M,t)$ of each 
shell, we ignore its effect on the radial motion 
of the shell during its infall. This is justified 
because a shell can only fall into a black hole if 
its angular momentum is extremely small. Fourth, we 
ignore the corrections to the radial motion of the 
shells due to the wave nature of the axion fluid. 
This is equivalent to describing the infalling
axion waves in the WKB (or eikonal) approximation,
and is tantamount to ignoring the role of so-called 
"quantum pressure".  It is justified to do so 
provided that the axion Compton wavelength is much 
less than the black hole size. For our smallest 
black holes, which have mass of order 
$10^6 M_\odot$, the axion mass must be 
more than of order $10^{-16}\,{\rm eV}/c^2$. 
The condition is amply satisfied by QCD axions. 

Eqs.~(\ref{radinf}) describe the evolution of an 
overdensity entirely in terms of the function 
$r_0(M)$ that gives the turnaround radius of each 
shell. We defined our initial time $t_{\rm in}$ 
to be the turnaround time of the innermost shells
in Eq.~(\ref{eq:t_in}) which can be alternatively 
expressed as
\[
  t_{\rm in} = \lim_{M \rightarrow 0} t_0(M) \,.
\]
At that time, the density near the center has the form 
\begin{equation}
  \rho(r,t_{\rm in}) = \rho(0,t_{\rm in})
  \left[1 - \left({r \over R}\right)^2 + \mathcal{O}\left(\frac{r^3}{R^3}\right) \right] \,.
\label{isden}
\end{equation}
We require $r_0(M)$ to be such that Eq.~(\ref{isden}) 
is reproduced for $r \ll R$, and such that 
$\rho(r,t_{\rm in})$ approaches the 
contemporary average cosmological energy 
density $\bar{\rho}(t_{\rm in})$ for $r \gg R$.
The requirements are met by the choice: 
\begin{equation}
  r_0(M) = R\left[\left({M \over M_f}\right)^{1 \over 3}
  + {1 \over 5}{M \over M_f} 
  + {1 \over 2} \left({M \over M_f}\right)^2\right]
\label{rom}
\end{equation}
with
\begin{equation}
  M_f = {4 \pi \over 3} \rho(0,t_{\rm in}) R^3 \,.
\label{mf}
\end{equation}
The resulting initial density profile $\rho(r,t_{\rm in})$
is shown in Fig.~2. 

The central overdensity collapses at time 
\begin{equation}
 t_{\rm coll} = 2 t_{\rm in} 
 = \pi \sqrt{3 \over 8 \pi G \rho(0,t_{\rm in})} \,,
\label{coll}
\end{equation}
with the outer shells collapsing later. We define 
$M_\star(t)$ such that all axions within shell 
$M_\star(t)$ (but outside any black hole that may have 
formed) thermalize sufficiently fast that they rotate 
rigidly, in the three-dimensional sense. Let 
$\omega(t)$ be their angular rotation frequency at 
that time. For $M < M_\star(t)$ the specific angular 
momentum of the axions at the equator of shell $M$ is 
therefore
\begin{equation}
  L(M,t) = \omega(t) r(M,t)^2 \,.
\label{angm}
\end{equation} 
Its rate of change following the motion is 
\begin{equation}
  \dot{L}(M,t) = {d \omega \over dt}(t) r(M,t)^2 
  + 2 \omega(t) r(M,t) v_r(M,t) \,,
\label{Ldot}
\end{equation} 
where $v_r(M,t) = \dot{r}(M,t)$ is the radial velocity 
of the shell. The maximum rate at which the shell can 
shed its specific angular momentum $L(M,t)$ while 
collapsing is given by Eq.~(\ref{maxLdot}) with 
$r = r(M,t)$.

If there were no relaxation, $\dot{L}(M,t) = 0$ 
and the angular frequency of each shell would 
increase as $r(M,t)^{-2}$. Instead, relaxation 
allows the shells within $M_\star(t)$ to collapse 
without hardly increasing their angular rotation 
frequency $\omega(t)$ because their angular momentum 
is transported outward. The condition for a shell 
to collapse without hardly increasing its angular 
rotation frequency is that the RHS of Eq.~(\ref{maxLdot}) 
is larger than the second term on the RHS of 
Eq.~(\ref{Ldot}), or equivalently that 
\begin{equation}
  \omega(t) \lesssim
  2 \pi G \rho(M,t) r(M,t) {1 \over v_r(M,t)} 
  \equiv \omega_{\rm max}(M,t) \,.
\label{omax}
\end{equation}
$M_\star(t)$ is therefore the largest shell such that 
\begin{equation}
  \omega(t) < \omega_{\rm max}(M, t)
\label{Mstarc}
\end{equation}
for all $M < M_\star(t)$. Because 
$\omega_{\rm max}(M,t)$ is, at all times, a decreasing 
function of $M$ near $M=0$, $M_\star(t)$ is the 
smallest solution of
\begin{equation}
  \omega_{\rm max}(M_\star(t),t) = \omega(t) \,.
\label{Mstar}
\end{equation}
All axions within the volume enclosed by $M_\star(t)$ 
exchange angular momentum sufficiently fast that they 
can, and therefore do, rotate with the common angular 
frequency $\omega(t)$. On the other hand all axions 
outside shell $M_\star(t)$ conserve their angular 
momentum. Of course, the transition at $M_\star(t)$ 
is not sudden as we take it to be but smoothing it 
out is not expected to change the final outcomes significantly.

$\omega(t)$ increases with time for two 
distinct reasons:
\begin{equation}
  {d \omega \over dt} = {d \omega \over dt}\Big|_L 
  + {d \omega \over dt}\Big|_T \,.
\label{omdot}
\end{equation}
The first term is due to the conservation of angular 
momentum within the volume of axions that rotate 
rigidly, and the second term is due to tidal torquing. 
The first term is 
\begin{equation}
  {d \omega \over dt}\Big|_L = - \omega(t) 
  {\dot{I}(t) \over I(t)}
\label{domL}
\end{equation}
with $I(t)$ the moment of inertia of all 
axions between shells $M_{\rm bh}(t)$ and 
$M_\star(t)$, where $M_{\rm bh}(t)$ is the 
black hole mass at time $t$,
\begin{equation}
  I(t) = {2 \over 3} \int_{M_{\rm bh}(t)}^{M_\star(t)}
  dM\,r(M,t)^2 \,,
\label{mom}
\end{equation}
and
\begin{equation}
  \dot{I}(t) = {4 \over 3} 
  \int_{M_{\rm bh}(t)}^{M_\star(t)}
  dM\,r(M,t) v_r(M,t)
\label{dmom}
\end{equation}
is its time derivative following the motion. 
The second term on the RHS of Eq.~(\ref{omdot})
is estimated in Appendix~B. Let us write the 
initial value of $\omega(t)$ as
\begin{equation}
  \omega(t_{\rm in}) \equiv \omega_{\rm in} 
  = {j \over t_{\rm in}} \,.
\label{omin}
\end{equation}
In case of rigid rotation, the relationship 
between $j$ and the spin parameter is
\begin{equation}
  \lambda = {4 \over 5 \pi} \sqrt{6 \over 5} j 
  = 0.279j + {\cal O}(j^2) \,.
\label{jom}
\end{equation}
We therefore expect $j$ to be in the approximate 
range of 0.03 to 0.8. We find in Appendix~B that
\begin{equation}
  {d \omega \over dt}\Big|_T \simeq 
  2.2j~{1 \over t_{\rm in}^2}
  \left({t_{\rm in} \over t}\right)^{4 \over 3}
\label{domT}
\end{equation}
during the interval $t_{\rm in} < t< t_{\rm coll}$.

Eqs.~(\ref{omax}), (\ref{Mstar}), (\ref{omdot}),
(\ref{domL}) and (\ref{domT}) were solved numerically. 
Initially, $M_\star$ is of order $M_f$ or larger. As 
$\omega(t)$ increases, it becomes more and more 
difficult to maintain rigid rotation, $M_\star(t)$ 
starts to decrease and then accelerates towards 
$M = 0$. Provided $\omega$ is sufficiently small at 
time $t_{\rm coll}$, a black hole forms and grows. 
Soon thereafter, at a time $t_f$, $r(M_\star(t),t)$ 
reaches zero and relaxation stops. After 
$t_f$, $L(M,t)$ of each shell is conserved. The black 
hole mass is the largest $M$ that satisfies 
$r_{\rm min}(M,t_f) < 2 GM/c^2$ or 
equivalently $L(M,t_f) < 2 G M/c$. All the shells that 
do not fall into the black hole move back out and start
to form the halo that will later surround the galaxy.
Although for the smallest expected $j$ values 
($j\sim 0.03$) a large fraction of the initial 
overdensity falls into the black hole, the total 
fraction of dark matter that ends up in supermassive 
black holes is very small since the initial 
overdensity is only a small fraction 
(of order $10^{-4}$) of the galaxy it grows into.

For given $j$ the black hole mass $M_{\rm bh}$ 
is found to be very nearly proportional to $M_f$.
It has only a small dependence on $t_{\rm in}$ 
for fixed $M_f$ and $j$. Fig.~3 shows 
$M_{\rm bh}$ as a function of $j$ for
a) $M_f = 3 \times 10^{10} M_\odot$,
b) $M_f = 10^{9} M_\odot$, and
c) $M_f = 3 \times 10^{7} M_\odot$. 
For the sake of definiteness, collapse was
assumed to occur at redshift $z = 10$, so 
that $t_{\rm coll} \simeq$ 0.5~Gyr, and 
hence $t_{\rm in} \simeq$ 0.25~Gyr, implying
$\rho(0,t_{\rm in}) \simeq 7.0 \times 10^{-26}$~gr/cc.
The three cases were chosen to correspond to 
overdensities that will evolve later into 
a) very large galaxies, b) galaxies similar to the 
Milky Way and c) small galaxies. Since the ratio 
$M_{\rm bh}/M_f$, a function of $j$, has little 
dependence on the collapse time, it is straightforward 
to generate results for all plausible cases.

Fig.~3 shows that no black hole forms for 
$j>j_c\sim 0.87$. It also implies that black holes of 
mass less than $10^{-5}$ of the initial overdensity 
are unlikely because this occurs only for a very 
small range of $j$ values, $0.85 < j < 0.87$. 
Furthermore, near the cutoff at $j=0.87$ the 
formation of a black hole would be easily disrupted 
by the various complications, such as the anisotropy 
of the infall, that we have neglected.

The black hole masses that result from the initial 
collapse of axion overdensities near cosmic dawn are 
in surprisingly good agreement with observations. 
First, the range of black hole masses formed, from 
approximately $10^5 M_\odot$ to a few times 
$10^{10} M_\odot$ is the mass range of observed 
supermassive black holes. As mentioned already, a 
cutoff is predicted at low masses. Specifically the 
theory predicts that a black hole mass less than 
approximately $10^5 M_\odot$ is unlikely on the 
scale of Milky Way size galaxies. Also, black hole 
masses larger than $10^{11} M_\odot$ are unlikely. 
They occur only in the largest overdensities and 
only if $\omega_{\rm in}$ is unexpectedly small. 
Second, the predicted supermassive black holes form 
near cosmic dawn. The puzzle of why supermassive 
black holes appear at high redshifts is removed. 
Although they may merge and accrete later, mergers 
and accretion are not necessary to explain their size. 
Third, there is a strong correlation between black 
hole mass and galaxy size. On the other hand, for a 
given galaxy size the black hole mass ranges widely, 
by a factor hundred or so. Both the correlation with 
galaxy size and the intrinsic variability are in 
qualitative agreement with observation. Fourth, the 
black holes form for approximately the range of 
galactic spin parameters expected from tidal torquing. 
Since there are galaxies with black holes close to 
the cutoff in $j$, e.g. the Milky Way whose central 
black hole has mass $4.3 \times 10^6 M_\odot$, the 
theory suggests that there are galaxies without 
supermassive black hole because their $j$ happens to 
be larger than the cutoff. This too appears 
consistent with observation. For example, the nearby 
small galaxy M33 appears not to have a supermassive 
black hole \cite{M33}.

Our description of supermassive black hole formation 
does not make any ad-hoc assumption. No new particle 
is postulated, other than the standard QCD axion or 
an axion-like particle with similar properties. The
invoked processes of thermalization and angular 
momentum transport in the cold axion dark matter 
fluid are the same as were described previously in 
Refs.~\cite{CABEC,Erken,Banik}. The theory is 
predictive and can be tested further. It should be 
possible to predict the distribution of supermassive 
black hole masses from the distribution of 
overdensities of a given size at cosmic dawn and 
the distribution of galactic spin parameters. The 
amplitude and spectrum of gravitational waves 
produced can be calculated, to be compared with
observation. By including baryons, it may be possible 
to determine whether the theory is consistent with 
the observed relation \cite{Msigma} between the 
stellar velocity dispersion in a galactic bulge and 
the mass of the supermassive black hole mass at its center.

\begin{acknowledgments}

We are grateful to Laura Blecha for critical 
feedback and to Jeff Andrews, Abhishek Chattaraj,
Jeff Dror, Antonios Kyriazis, Wei Xue and Fengwei 
Yang for useful discussions. This work was supported 
in part by the U.S. Department of Energy under 
grant DE-SC0022148 at the University of Florida. 

\end{acknowledgments}

\newpage 

\appendix
\section{Axion fluid fluctuations}

It has been shown in a variety of contexts that 
degenerate Bosonic systems have generically large 
fluctuations in intensity or density \cite{Pathria}. 
In this appendix we show this for the cold dark 
matter axion fluid.

In the non-relativistc limit, the scalar 
quantum field $\phi(\vec{x},t)$ describing 
axions in a volume $V$ may be written as
\begin{equation}
  \phi(\vec{x},t) = {1 \over \sqrt{2m}}
  [\psi(\vec{x},t) e^{- i m t} + \mathrm{h.c.}]
\label{axf}
\end{equation}
and expanded 
\begin{equation}
  \psi(\vec{x},t) = \sum_{\vec{\alpha}}
  u^{\vec{\alpha}}(\vec{x},t) a_{\vec{\alpha}}(t)
\label{exp}
\end{equation} 
where the $u^{\vec{\alpha}}(\vec{x},t)$
are any set of orthonormal and complete (ONC) 
wavefunctions in that volume:
\begin{equation}
  \int_V d^3x\, u^{\vec{\alpha}}(\vec{x},t)^*
  u^{\vec{\beta}}(\vec{x},t) = 
  \delta^{\vec{\alpha}\vec{\beta}}
  \,,\quad
  \sum_{\vec{\alpha}} u^{\vec{\alpha}}(\vec{x},t)
  u^{\vec{\alpha}}(\vec{y},t)^* = 
  \delta^{(3)}(\vec{x} - \vec{y}) \,.
\label{onc}
\end{equation}
The $a_{\vec{\alpha}}(t)$ and their Hermitian 
conjugates $a_{\vec{\alpha}}(t)^\dagger$
satisfy canonical equal time commutation 
relations. The most general axion system 
state is given by a linear combination
\begin{equation}
  |c_{\{\cal N\}} \rangle = \sum_{\{\cal N\}}
  c_{\{\cal N\}} |\{\cal N\}\rangle
\label{genst}
\end{equation}
of all possible particle state occupation number 
eigenstates
\begin{equation}
  |\{{\cal N}\}\rangle = \prod_{\vec{\alpha}}
  {1 \over \sqrt{{\cal N}_{\vec{\alpha}} !}}
  (a_{\vec{\alpha}}^\dagger)^{{\cal N}_{\vec{\alpha}}} 
  |0\rangle \,.
\label{pnst}
\end{equation}
Here $|0\rangle$ is the empty state and
$\{{\cal N}\} = \{{\cal N}_{\vec{\alpha}}:
\forall \vec{\alpha}\}$ is the set of integers 
giving the occupation number of each particle state. 
$N = \sum_{\vec{\alpha}} {\cal N}_{\vec{\alpha}}$
is the total number of axions in volume $V$.

For the purpose of describing the cold dark 
matter axion fluid with average number density 
${1 \over m} \rho(\vec{x},t)$ and average 
velocity field $\vec{v}(\vec{x},t)$ we choose 
a set of ONC wavefunctions as the wavefucntion 
of Eq.~(\ref{psi}) and spatial modulations thereof 
with wavevector $\vec{k}$:
\begin{equation}
  u^{\vec{k}}(\vec{x},t) =
  e^{i \vec{k} \cdot \vec{\chi}(\vec{x},t)} \Psi(\vec{x},t)
  \sim e^{i \vec{k} \cdot \vec{x}} \Psi(\vec{x},t) \,. 
\label{sonc}
\end{equation}
If the axion fluid is homogeneous and at rest, 
this ONC set of wavefunctions would simply be
\begin{equation}
  u^{\vec{k}}(\vec{x}) = {1 \over \sqrt{V}} 
  e^{i \vec{k} \cdot \vec{x}} \,.
\label{simp}
\end{equation}
When the axion fluid is inhomogeneous and/or moving, 
there are many ways to construct suitable 
$u^{\vec{k}}(\vec{x},t)$. A particular method is 
described in Ref.~\cite{Chak}. Another way is to 
start with all the wavefuntions 
$e^{i \vec{k} \cdot \vec{x}} \Psi(\vec{x},t)$ and 
orthonormalize them in succession. In any such 
basis, the state of the axion fluid is one where 
most particle states have low occupation numbers 
but those with wavevector magnitude 
$k \lesssim m \delta v/\hbar$ are hugely occupied. 
The occupation numbers of those states that are 
occupied are of order $10^{61}$ or larger \cite{CABEC,Erken}. 

The number density of axions is the operator
\begin{equation}
  n(\vec{x},t) = \psi(\vec{x},t)^\dagger \psi(\vec{x},t) \,.
\label{denop}
\end{equation}
In occupation number eigenstates, it has quantum 
mechanical average 
\begin{equation}
  \langle \{{\cal N}\}| n(\vec{x}, t) |\{{\cal N}\} \rangle
  = \sum_{\vec{k}} {\cal N}_{\vec{k}} |u^{\vec{k}}(\vec{x},t)|^2
  = {1 \over m} \rho(\vec{x},t) \,.
\label{avden}
\end{equation}
Let us define the operator
\begin{equation}
  \delta n(\vec{x},t) = 
  n(\vec{x},t) - {1 \over m} \rho(\vec{x},t) \,,
\label{delden}
\end{equation}
one readily finds 
\begin{equation}
  \langle \{{\cal N}\}| \delta n(\vec{x},t) 
  \delta n(\vec{y},t) | \{{\cal N}\} \rangle =
  |D(\vec{x}, \vec{y}; t)|^2~[
  1 + {\cal O}\left({1 \over {\cal N}}\right)] \,,
\label{corden}
\end{equation}
where 
\begin{equation}
  D(\vec{x}, \vec{y}; t) = 
  \sum_{\vec{k}} {\cal N}_{\vec{k}}~ 
  u^{\vec{k}}(\vec{x},t)^* u^{\vec{k}}(\vec{y},t) \,.
\label{corfct}
\end{equation}
We have therefore
\begin{equation}
  \langle \{{\cal N}\}| (\delta n(\vec{x},t))^2
  \{{\cal N}\} \rangle =
  \left({1 \over m} \rho(\vec{x}, t) \right)^2
  \left[1 + {\cal O}\left({1 \over {\cal N}}\right)\right] \,.
\label{corden2}
\end{equation}
Thus, in eigenstates of the occupation numbers, 
the root-mean-square deviation from the average 
density at every space-time point equals the 
average density there. Moreover, these 
deviations are correlated over distances 
of order $\ell = {1 \over \delta k} = 
{\hbar \over m \delta v}$ since $D(\vec{x}, \vec{y}; t)$
cannot vary much over distances shorter than $1/\delta k$. 

In the general system states of Eq.~(\ref{genst})
it is not possible to make such strong statements
as Eq~(\ref{corden2}) because they include system 
states in which all axions are in a single state
whose wavefunction is a linear combination of 
several $u^{\vec{k}}(\vec{x},t)$. In these
very special system states, the quantum mechanical 
uncertainty in measuring $n(\vec{x},t)$ vanishes.
Consider nonetheless the quantum mechanical average
in a general state of any operator $\Omega$
\begin{equation}
  \langle c_{\{{\cal N}\}} | \Omega | c_{\{{\cal N}\}} \rangle
  = \sum_{\{{\cal N}\}} \sum_{\{{\cal N}^\prime\}}
  c_{\{\cal N}\}^* c_{\{{\cal N}^\prime\}} 
  \langle \{{\cal N}\} | \Omega | \{{\cal N}^\prime\}  \rangle \,.
\label{genav}
\end{equation}
Even in case the initial values of the complex 
coefficients $c_{\{{\cal N}\}}$ are very special, 
they acquire random phases after some time, 
especially as a result of thermal relaxation 
of the fluid. We have 
\begin{equation} 
  \cdot \langle c_{\{\cal N\}} | \Omega | c_{\{{\cal N}\}} \rangle \cdot
  = \sum_{\{{\cal N}\}} |c_{\{\cal N\}}|^2 
  \langle \{{\cal N}\} | \Omega | \{{\cal N}\} \rangle \,,
\label{rand}
\end{equation}
where $~~\cdot \langle \dots \rangle \cdot~~$ means 
quantum mechanical average followed by an average 
over the phases of the coefficient $c_{\{{\cal N}\}}$.
Using Eq.~(\ref{corden2}) we have 
\begin{equation}
  \cdot \langle c_{\{{\cal N}\}}|
  (\delta n(\vec{x}, t))^2 | c_{\{{\cal N}\}} \rangle \cdot
  = \left({1 \over m}\rho(\vec{x}, t)\right)^2 \,.
\label{fin}
\end{equation} 

That degenerate Bose fluids generically have 
large fluctuations is not a new result.  In 
a 1909 paper A. Einstein showed that the 
fluctuations in Bose fluids are not Poisson 
distributed and that the particles in such 
fluids tend to be more bunched than classical 
particles.  The phenomenon has been thoroughly 
investigated, theoretically and experimentally, 
in the case of light beams.  For a review, see 
ref. \cite{Mandel}. In particular, E.W. Purcell 
\cite{Purcell} showed that highly degenerate 
photon beams  have $\delta \rho = \rho$ 
fluctuations correlated over a time scale 
of order the inverse of the photon frequency 
dispersion.  The phenomenon can be seen also 
in a classical field theory description of 
degenerate Bose fluids.  It is straightforward 
to show that a linear combination of many 
plane waves has $\delta \rho = \rho$ density 
fluctuations if the wave amplitudes are 
Gaussian distributed and their phases are 
random.   

As mentioned already, there are special system 
states in which the fluctuations are absent.  
In particular $\delta \rho = 0$ if all the 
particles are in a single state. (A photon 
beam stops fluctuating in time if its frequency 
dispersion vanishes, but of course a perfectly
monochromatic photon beam is an idealization.) 
Now, Bose-Einstein condensation does move a 
large fraction of all particles into a single 
state, to wit the lowest energy available state.  
If the condensate is stable and its environment 
is static, the fluctuations disappear when 
the Bose-Einstein condensation is complete.  
However, this does not apply to the axion 
fluid under consideration here  because it 
is collapsing by gravitational instability.
The Bose-Einstein condensation of the axions 
is never complete because the lowest energy 
particle state keeps changing.  Rigid 
rotation can only be maintained if the 
axions keep moving between states of 
different angular momentum.  In the 
collapsing overdensity numerous axion 
states have enormous occupancy which, 
as shown above, is the condition for the 
$\delta\rho = \rho$ fluctuations to be 
present.

\newpage 

\section{Tidal torquing of the overdensity}

Consider the tidal torque on all the particles
between shells $M_{\rm bh}(t)$ and $M_\star(t)$
\begin{equation}
  \vec{\tau}(t) = \int d\Omega
  \int_{r(M_{\rm bh}(t),t)}^{r(M_\star(t),t)} r^2 dr\,
  \rho(\vec{r},t)~ \vec{r} \times
  ( - \vec\nabla \Phi(\vec{r},t)
  + \vec\nabla \Phi(\vec{0}, t)) \,,
\label{domT2}
\end{equation}
where $\Phi(\vec{r},t)$ is the gravitational
potential due to density perturbations outside 
the overdensity of interest. The region 
surrounding the overdensity near cosmic dawn 
is an expanding Einstein-de Sitter space-time 
where $\vec{r} = \vec{x}
\left({t \over t_{\rm in}}\right)^{2 \over 3}$,
$\vec{x}$ are comoving coordinates, and
$\Phi(\vec{r}, t) = \Phi(\vec{x})$ \cite{SW}.
We expand the gravitational potential in Taylor
series about the center of the overdensity
\begin{equation}
  \Phi(\vec{x}) = \Phi(\vec{0}) - \vec{g}\cdot\vec{x}
  + {1 \over 2} \vec{x}^T \mathbb{T} \vec{x} + \dots \,,
\label{PniT}
\end{equation}
where $\vec{g}$ is a constant vector and
$\mathbb{T}$ a constant symmetric matrix.
Eq.~(\ref{domT2}) becomes
\begin{equation}
  \vec{\tau}(t) \simeq -
  \left({t \over t_{\rm in}}\right)^{4 \over 3}
  \int d\Omega
  \int_{r(M_{\rm bh}(t),t)}^{r(M_\star(t),t)} r^4 dr\,
  \rho(\vec{r},t)~
  \hat{n}(\theta, \phi) \times \mathbb{T}~\hat{n}(\theta,\phi) \,,
\label{domT3}
\end{equation}
where $\hat{n}(\theta,\phi) = {1 \over r} \vec{r}$ is the unit 
vector in the direction of spherical coordinates $(\theta,\phi)$.
We expect the asphericity of the smoothed overdensity, as in 
Eq.~(\ref{inden}), to dominate the integral in Eq.~(\ref{domT3}). 
Let us rewrite Eq.~(\ref{inden}) as
\begin{equation}
  \rho(\vec{r}, t_{\rm in}) = 
  \rho(\vec{0}, t_{\rm in})
  \left[1 - \left({r \over R}\right)^2 A(\theta,\phi)\right] \,,
\label{inden2}
\end{equation}
where $A(\theta,\phi)$ is a linear combination 
of the spherical harmonics up to second order.

For $0 < t < t_{\rm in}$, we estimate the 
integral in Eq.~(\ref{domT3}) using linear 
perturbation theory in a homogeneous expanding 
universe of density $\rho(\vec{0},t)$. In such 
a universe
\begin{equation}
  r(M,t) = b(t) r(M,t_{\rm in}) \,,
\label{bexp}
\end{equation}
where the scale factor $b(t)$ is given by 
\begin{eqnarray}
  b(t) &=& \sin^2 \sigma
  \nonumber\\
  t &=& \sqrt{3 \over 8 \pi G \rho(\vec{0}, t_{\rm in})}
  [\sigma - {1 \over 2} \sin(2 \sigma)] \,.
\label{bsf}
\end{eqnarray}
We have therefore 
\begin{equation}
  {\delta \rho(r(M,t) \hat{n}, t) \over \rho(\vec{0}, t)}
  = \delta_+(t) 
  {\delta \rho(r(M,t_{\rm in}) \hat{n}, t_{\rm in}) 
  \over \rho(\vec{0}, t_{\rm in})} \,,
\label{linpt}
\end{equation}
where $\delta_+(t)$ is the appropriate growth function 
\cite{SW} normalized so that $\delta_+(t_{\rm in}) = 1$:
\begin{equation}
  \delta_+(\sigma) = {1 \over 2}
  \left({ - 3 \sigma \cos \sigma \over \sin^3 \sigma}
  + {3 \over \sin^2 \sigma} - 1\right) \,.
\label{dplus}
\end{equation}
Since
\begin{equation}
  \delta \rho(r(M,t_{\rm in}) \hat{n}, t_{\rm in}) = 
  - \rho(\vec{0}, t_{\rm in}) 
  \left({r(M, t_{\rm in}) \over R}\right)^2 
  A(\theta,\phi)
\label{drin}
\end{equation}
we have 
\begin{equation}
  \delta \rho(r(M,t) \hat{n}, t) = - \delta_+(t)
  \rho(\vec{0}, t) \left({r(M, t) \over R}\right)^2 
  A(\theta, \phi) \,.
\label{drt}
\end{equation}
We substitute this for $\rho(\vec{r},t)$ in 
Eq.~(\ref{domT3}) and define 
\begin{equation}
  {3 \over 8 \pi} \int d\Omega A(\theta,\phi)
  \hat{n}(\theta,\phi) \times \mathbb{T}~\hat{n}(\theta,\phi)
  \equiv \Sigma \hat{z} \,.
\label{bsig}
\end{equation}
Upon integrating the radial coordinate $r$ from 
0 to $R(t) = b(t) R$, Eq.~(\ref{domT3}) becomes
\begin{equation}
  \vec{\tau}(t) \simeq {8 \pi \over 3} \Sigma \hat{z}
  \left({t_{\rm in} \over t}\right)^{4 \over 3} 
  {1 \over 7} R^5 b(t)^7 \rho(\vec{0},t) \delta_+(t) \,.
\label{domT4}
\end{equation} 
Since
\begin{equation}
  \vec{\tau}(t) = I(t) \hat{z} {d \omega \over dt}\Bigg|_T
  \label{obv}
\end{equation}
with
\begin{equation}
  I(t) = {2 \over 3} \int_0^{R(t)} 4 \pi r^4 dr~\rho(\vec{0},t)
  = 4 \pi {2 \over 15} R^5 b(t)^5 \rho(\vec{0},t) \,,
\label{moml}
\end{equation}
we have 
\begin{equation}
  {d \omega \over dt}\Bigg|_T \simeq {5 \over 7}
  \Sigma \left({t_{\rm in} \over t}\right)^{4 \over 3}
  b(t)^2 \delta_+(t) 
\label{domT5}
\end{equation}
during the time interval $0 < t < t_{\rm in}$. Hence 
\begin{equation}
  \omega(t_{\rm in}) \simeq {5 \over 7} \Sigma 
  \int_0^{t_{\rm in}} dt 
  \left({t_{\rm in} \over t}\right)^{4 \over 3}
  b(t)^2 \delta_+(t)
  \simeq 0.448~\Sigma~t_{\rm in} \,.
\label{omin3}
\end{equation}

During the time interval $t_{\rm in} < t < t_{\rm coll}$, 
tidal torquing continues but perturbation theory breaks 
down. The mass distribution becomes dominated by its
anisotropic component. We estimate the integral in 
Eq.~(\ref{domT3}) by setting 
\begin{equation}
  \rho(r(M,t) \hat{n}(\theta,\phi),t) = \rho(M,t) A(\theta,\phi) \,,
\label{estim}
\end{equation}
which yields
\begin{equation}
  \vec{\tau}(t) \simeq \left({t_{\rm in} \over t}\right)^{4 \over 3}
\Sigma~I(t)~\hat{z}
\label{tau4}
\end{equation}
and hence 
\begin{equation}
  {d \omega \over dt}\Bigg|_T \simeq 
  \left({t_{\rm in} \over t}\right)^{4 \over 3} \Sigma \,.
\label{domT7}
\end{equation}
Combining this with Eq.~(\ref{omin3}) yields
\begin{equation}
  {d \omega \over dt} \Bigg|_T \simeq 2.2~
  {\omega(t_{\rm in}) \over t_{\rm in}} 
  \left({t_{\rm in} \over t}\right)^{4 \over 3}
\label{domT8}
\end{equation}
during the time interval $t_{\rm in} < t < t_{\rm coll}$.

\newpage

\maxdeadcycles=200

\begin{figure}
\begin{center}
\includegraphics[height=110mm]{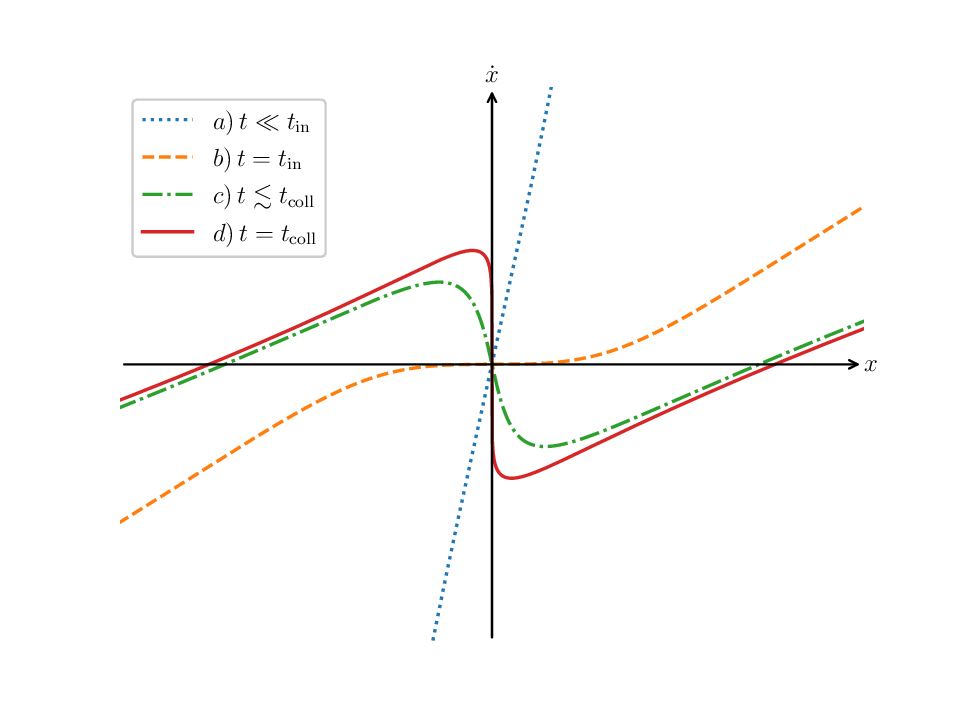}
\vspace{0.3in}
\caption{Phase space distribution of cold collisionless
particles during the collapse of a large smooth overdensity 
near cosmic dawn, at four different times: a) just after 
the Big Bang, b) when the central part of the overdensity 
is at turnaround, c) just before, and d) at the time 
$t_{\rm coll}$ of collapse of the central overdensity. 
$x$ is the spatial coordinate along an arbitrary direction 
through the overdensity. An actual overdensity has small 
scale structure which has been smoothed out in the figure.}
\end{center}
\label{fig:phsp}
\end{figure}

\begin{figure}
\begin{center}
\includegraphics[height=110mm]{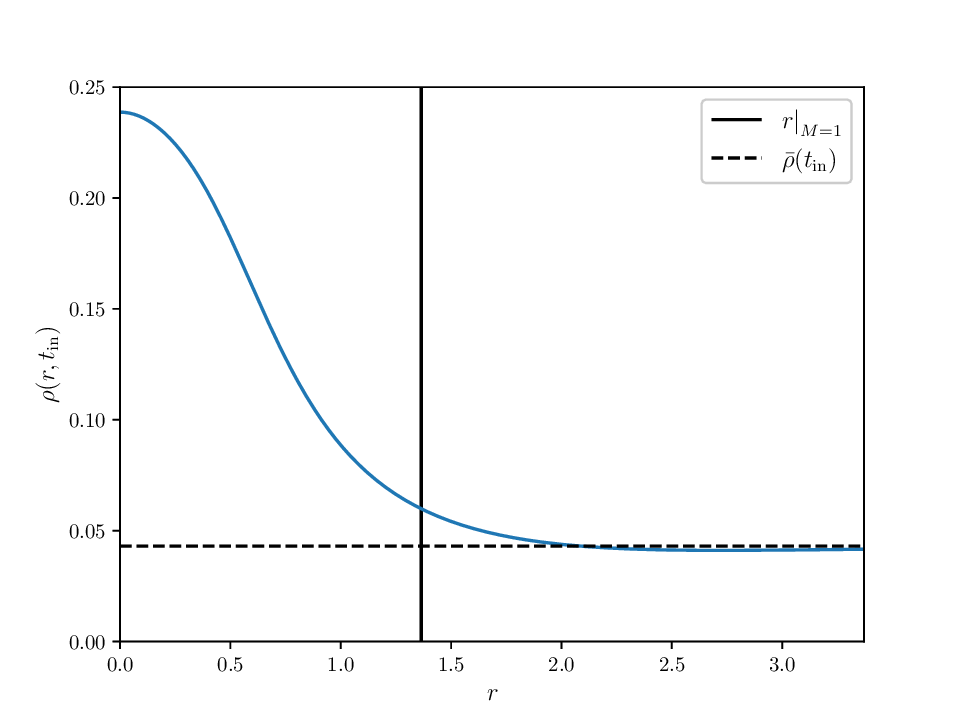}
\vspace{0.3in}
\caption{Density profile at time $t_{\rm in}$, when the 
central part of the overdensity is at turnaround, in 
units where $M_f = 1$ and $R = 1$. In these units 
the central density $\rho(\vec{0}, t_{\rm in}) = 3/4 \pi$ 
and the contemporary average cosmological energy density 
$\bar{\rho}(t_{\rm in}) = 4/3 \pi^3$. The average cosmological 
energy density is indicated by the horizontal dashed line. 
The vertical solid line indicates the radius that contains 
mass $M_f$.}
\end{center}
\label{fig:profile}
\end{figure}

\begin{figure}
\begin{center}
\includegraphics[height=110mm]{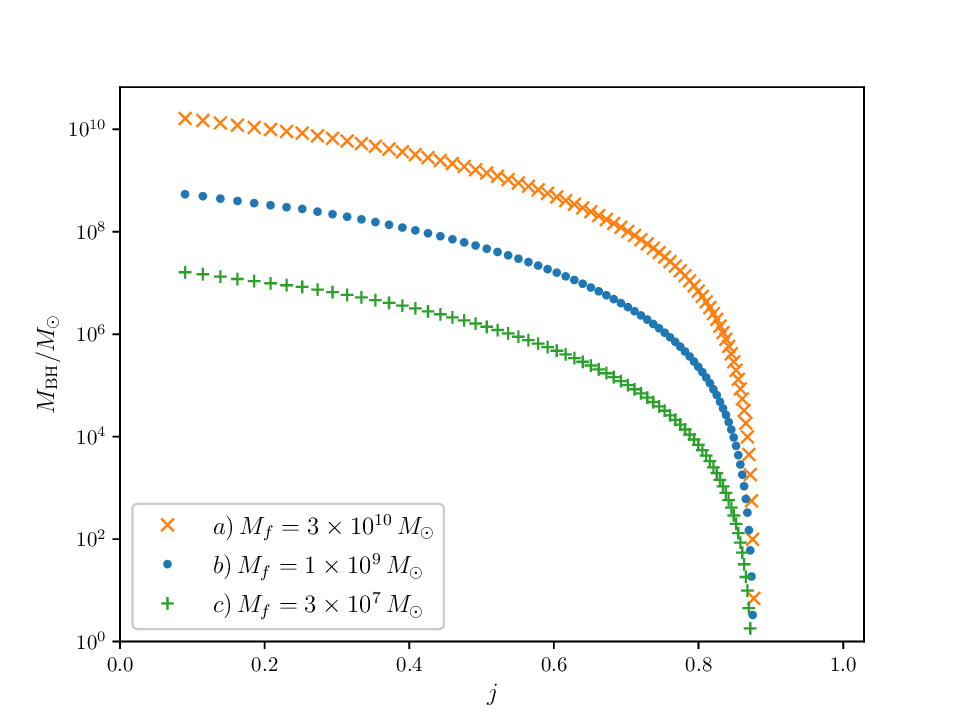}
\vspace{0.3in}
\caption{Black hole mass as a function of 
$j \equiv \omega_{\rm in} t_{\rm in}$ for 
three values of $M_f$. For given $M_f$
there is only a very slight dependence 
of the black hole mass on $t_{\rm in}$. 
The values shown were computed for 
$z_{\rm coll}$ = 10 and hence 
$t_{\rm in}$ = 240~Myr.}
\end{center}
\label{fig:mbh}
\end{figure}

\clearpage


\newpage

\begin{thebibliography}{bib}

\bibitem{Korm}
J. Kormendy and L.C. Ho, Ann. Rev. of Astron.
and Astroph., 51 (2013) 511.

\bibitem{EHT}
A. Kazunori et al. (the EHT Collaboration), 
Ap. J. Lett. 875 (2019) L5, and Ap. J. Lett. 
930 (2022) L12.

\bibitem{AGN}
J. Frank, A. King and D.J. Raine,
{\it Accretion Power in Astrophysics},
Third Edition, Cambridge University 
Press, 2002.

\bibitem{PTA}
G. Agazie et al. (the NANOGrav. Collaboration), 
Ap. J. Lett. 951 (2023) L8;
D.J. Reardon et al., Ap. J. Lett. 951 (2023) L6;
H. Xu et al., Res. Astron. Astroph. 23 (2023) 075024; 
J. Antoniadis et al. (the EPTA Collaboration), 
arXiv: 2306.16227.

\bibitem{bhfrev}
K. Inayoshi, E. Visbal and Z. Haiman, 
Ann. Rev. of Astron. and Astroph.,
58 (2020) 27.

\bibitem{SS}
S. Balberg and S.L. Shapiro, Phys. Rev. Lett. 
88 (2002) 101301; J. Pollack, D.N. Spergel
and P. Steinhardt, Ap. J. 804 (2015) 2, 131;
W.-X. Feng, H.-B. Yu and Y.-M. Zhong, Ap. J. 
Lett. 914 (2021) 2, L26.

\bibitem{KF}
T. Rindler-Daller, K. Freese, M.H. Montgomery, 
D. Winget and B. Paxton, Ap. J. 799 (2015) 210.

\bibitem{JWST}
R. Larson et al., Ap. J. Lett. 953 (2023) L29;
A. Bogdan et al., Nature Atron. 8 (2024) 126;
R. Maiolino et al., arXiv:2308.01230;
L.J. Furtak et al., arXiv: 2308.05735;
R. Miaolino et al., Nature 627 (2024) 59;

\bibitem{HoSt}
D. Hooper, A. Ireland, G. Krujaic
and A Stebbins, JCAP 04 (2024) 021.

\bibitem{CABEC}
P. Sikivie and Q. Yang, Phys. Rev. Lett. 
103 (2009) 111301.

\bibitem{Erken}
O. Erken, P. Sikivie, H. Tam and Q. Yang, 
Phys. Rev. D 85 (2012) 063520.

\bibitem{Zel}
M.Y. Khlopov, B.A. Malomed and Y.B. Zel'dovich, 
MNRAS 215 (1985) 575.

\bibitem{axdm}
J. Preskill, F. Wilczek and M. Wise, Phys. Lett.
B120 (1983) 127; L. Abbott and P. Sikivie,
Phys. Lett. B120 (1983) 133; M. Dine and W. Fischler,
Phys. Lett. B120 (1983) 137.

\bibitem{PoS}
P. Sikivie and Y. Zhao, arXiv:2601.xxxxx

\bibitem{Banik}
N. Banik and P. Sikivie, Phys. Rev. D 88 (2013) 123517.

\bibitem{Bard}
J.M. Bardeen, J.R. Bond, N. Kaiser and A.S. Szalay, 
Ap. J. 304 (1986) 15.

\bibitem{Peeb}
P.J.E. Peebles, Ap. J. 155 (1969) 393.

\bibitem{Efst}
G. Efstatathiou and B.J.T. Jones, MNRAS 186 (1979) 133;
J. Barnes and G. Efstathiou, Ap. J. 319 (1987) 575.

\bibitem{Hern}
X. Hernanadez, C. Park, B. Cervantes-Sodi and 
Y.-Y. Choi, MNRAS 375 (2007) 163.

\bibitem{anis}
C.C. Lin, L. Meistel and F.H. Shu,
Ap. J. 142 (1965) 1431;
Y.B. Zel'dovich, Astron. and Astroph. 5 (1970) 84;
J. Binney, Ap. J. 215 (1977) 492.

\bibitem{M33}
K. Gebhardt et al., Astron. J. 122 (2001) 2469.

\bibitem{Msigma}
F. Ferrarese and D. Merritt, Ap. J. 538 (2000) L9;
K. Gebhardt et al., Ap. J. (2000) L13.

\bibitem{Pathria}
R.K. Pathria and P.D. Beale, 
{\it Statistical Mechanics}, 3rd edition, 
Elsevier 2011, and references therein.

\bibitem{Chak}
S. Chakrabarty et al., Phys. Rev. D 97 (2018) 043531.

\bibitem{Einstein}
A. Einstein, Phys. Zeit. 10 (1909) 185.

\bibitem{Mandel}
L. Mandel, Progress in Optics 2 (1963) 181.

\bibitem{Purcell}
E.W. Purcell, Nature 178 (1956) 1449.

\bibitem{SW}
S. Weinberg, {\it Gravitation and Cosmology},
J. Wiley and Sons, 1972.

\end{thebibliography}
\end{document}